\documentclass[aps,pre,twocolumn]{revtex4-1}

\usepackage{amsmath}
\usepackage{amssymb}
\usepackage{color}

\usepackage{graphicx}

\DeclareMathAlphabet{\mathitbf}{OML}{cmm}{b}{it}

\newcommand{\dbar}{{\,\mathchar'26\mkern-12mu d}}

\begin{document}
\title{Unjamming in models with analytic pairwise potentials}
\author{Stefan Kooij$^{1,2}$ and Edan Lerner$^{1}$}
\affiliation{$^1$Institute for Theoretical Physics, University of Amsterdam, Science Park 904, Amsterdam, Netherlands\\ $^2$Van der Waals-Zeeman Institute, University of Amsterdam, Science Park 904, Amsterdam, Netherlands}

\begin{abstract}

The canonical models for studying the unjamming scenario in systems of soft repulsive particles assume pairwise potentials with a sharp cut-off in the interaction range. The sharp cut-off renders the potential non-analytic, but makes it possible to describe many properties of the solid in terms of the coordination number $z$, which has an unambiguous definition in these cases. Pairwise potentials without a sharp cut-off in the interaction range have not been considered in this context, but are of interest for understanding the relevance of the unjamming phenomenology to systems in which such a cut-off cannot be assumed. In this work we explore two systems with such interactions: an inverse power law and an exponentially decaying pairwise potential, with the control parameters being the exponent (of the inverse power-law) for the former and the number density for the latter. Both systems are shown to exhibit the characteristic features of the unjamming transition, among which are the vanishing of the shear to bulk modulus ratio and the emergence of an excess of low-frequency vibrational modes. We establish a relation between the hydrostatic pressure to bulk modulus ratio and the distance to unjamming in each of our model systems. This allows us to predict the dependence of other key observables on the distance to unjamming. Our results provide the means for a quantitative estimation of the proximity of generic glass forming models to the unjamming transition in the absence of a clear-cut definition of the coordination number, and highlight the general irrelevance of nonaffine contributions to the bulk modulus.  %   Our key results are that for the exponential interactions unjamming only occurs in the limit of vanishing density, and that unjamming occurs in the exponential case Using simple arguments, a mapping can be made from the control parameters of the canonical soft-sphere systems to the control parameters of the exponential and ipl systems, which allows us to predict the observed scaling laws. This implies that even though the contact number is ill-defined, one can still assign an effective contact number.

\end{abstract}

\maketitle

\section{introduction}
The unjamming scenario describes the abrupt loss of solidity of gently compressed soft particles or of elastic networks, that occurs when the coordination number $z$ is reduced towards the isostatic point $z_c = 2\dbar$, where $\dbar$ is the spatial dimension. This is typically achieved in soft spheres/discs by reducing the packing fraction $\phi$ towards the random close packing fraction $\phi_c$, or in elastic networks by removing interactions from the network. It is now well established that approaching the unjamming point is accompanied by the emergence of an excess of low-frequencies vibrational modes \cite{ohern2003,matthieu_thesis}, diverging correlation \cite{silbert2005} and response \cite{breakdown} lengthscales, and the vanishing of elastic moduli \cite{wouter_epl_2009}. Substantial attention was drawn by the unjamming scenario following proposals that it can explain the origin of several elusive glassy phenomena, such as the occurrence of the Boson Peak in glassy solids \cite{liu_boson_peak_prl,eric_emt_prestress} and the fiercely-debated fragility of supercooled liquids \cite{le_fragility}. 

Many of the interesting phenomena associated with the unjamming transition are adequately explained by variational \cite{matthieu_le_epl} and marginal stability \cite{matthieu_PRE_2005,nonlinear_excitations} arguments, and mean-field theories \cite{mw_EM_epl,Lubensky_CPA,phonon_gap,eric_emt_prestress}. A common theme to these approaches is the underlying assumption that pairs of the constituent particles or degrees of freedom (DOF) either interact or do not. In numerical investigations, this assumption is embodied by the specific form of pairwise interaction potentials employed in the canonical models; these are typically of the form
\begin{equation}
\varphi(r) \propto  \left\{\begin{array}{cc}(r-\ell)^\alpha\,,&r \le \ell\\0\,,&r > \ell\end{array}\right.\,,
\end{equation}
where $r$ is the distance between the centers of a pair of spherical particles, $\ell$ is the sum of their radii, and $\alpha$ is typically chosen to be 2 (harmonic interactions) or 5/2 (Hertzian interactions). This potential possesses a sharp cutoff at $r = \ell$, which leads to discontinuities in observables that depend on high order derivatives. For instance, it is well known that harmonic sphere packings posses a finite bulk modulus at the unjamming point \cite{ohern2003,matthieu_thesis}. While this is a non-trivial observation (e.g.~all elastic moduli in homogeneous random spring networks vanishes at the isostatic point \cite{wouter_epl_2009}), it would be impossible to observe this discontinuity if the pairwise interaction were analytic. This discussion highlights the importance of the general question: can unjamming occur if the constituent particles of a system interact via \emph{analytic} repulsive pairwise potentials, i.e.~potentials that do not possess a sharp cut-off in their interaction range? 

In this work we address this question by studying two different model systems of repulsive particles in two dimensions (2D), that can be driven to the unjamming point by tuning the appropriate control parameter. In the first system, particles interact via a potential that decays \emph{exponentially} with $r$ (EXP). We find the surprising result that this system only unjams in the limit of vanishing density, and not at a finite density as the canonical models do. We also study a constant-volume system of point-like particles interacting via an inverse power law $\varphi \propto r^{-\beta}$ (IPL), which is shown to unjam in the limit $\beta \to \infty$. We monitor a set of key observables as unjamming is approached: the ratio of shear to bulk moduli, the density of states and characteristic vibrational frequencies. We further explain their measured scaling laws using the unjamming framework, and the properties of the pairwise potentials employed. 

This work is organized as follows; in Sect.~\ref{numerics} we provide details of the models investigated and of the numerical methods employed throughout our work. Sect.~\ref{results} describes the unjamming phenomenology observed in our model systems. In Sect.~\ref{scaling_argument} we provide arguments that explain the scaling laws observed approaching the unjamming point. Our work is summarized in Sect.~\ref{summary}.

\section{Models and Methods}
\label{numerics}

As mentioned above, we employ two different models of repulsive particles in 2D. In this section we spell out the details of these models, and further discuss how key observables of interest are calculated. We end this section with an important discussion regarding the cutoff we introduced in the interaction range of the pairwise potentials, and its role in the observed phenomena. 

\subsection{The exponential model}
\begin{figure}[!ht]
\centering
\includegraphics[width = 0.50\textwidth]{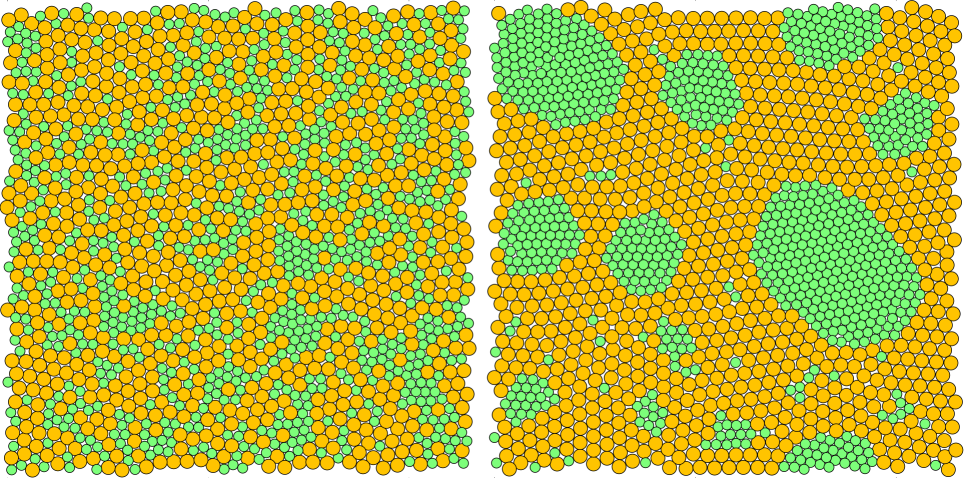}
\caption{\footnotesize Solid realizations of the EXP model. Obtaining a robust disordered solid depends on delicately tuning the model parameters. We show that changing the small-large interaction strength $\varepsilon_{ij}$ by a bit more than a percent, from 1.7 (left panel) to 1.72 (right panel) destabilizes the glass and leads to phase separation.}
\label{phase_separation_fig}
\end{figure}

The first model, referred to as the exponential model (or EXP in short), is a 50:50 binary mixture of `large' and `small' particles interacting via the pairwise potential
\begin{equation}
\varphi_{ij} =  \left\{\begin{array}{cc}\varepsilon_{ij}\left(e^{-r_{ij}/l_{ij}} + a_{ij}r_{ij} + b_{ij}\right)\,,&r \le r_c\\0\,,&r > r_c\end{array}\right.\,,
\end{equation}
where the constants $a_{ij},b_{ij}$ are determined such that the potential and its first derivative vanish at a cutoff distance $r_c$, which was set separately for each density to be larger than the second coordination shell (see discussion at the end of this Section). We discuss the importance of this cutoff and its role in the observed phenomena at the end of this Section. The interaction strengths $\varepsilon_{ij}$ are set to be $\varepsilon_0, 1.64\varepsilon_0$ and $3.05\varepsilon_0$ for a small-small, small-large or large-large interactions, respectively, where $\varepsilon_0$ is our microscopic units of energy. The interaction lengths $l_{ij}$ are set to be $l_0, 1.2l_0$, and $1.4l_0$ for a small-small, small-large, and large-large interactions, respectively, where $l_0$ is our microscopic unit of length. The glass-forming ability of the EXP system is very sensitive to the particular choice of these parameters, as demonstrated In Fig.~\ref{phase_separation_fig} and in \cite{youtube}.

%%%%%%%%%%%%%%%%%%%%%%%%%%%%%%%%%%%%%%%%%%%%%%%%%%%%%%%
%%%%%%%%%%%%%%%%%%%%%%%%%%%%%%%%%%%%%%%%%%%%%%%%%%%%%%

The key control parameter in the EXP model was the dimensionless density $\rho \equiv Nl_0^2/V$, which was varied between $5.6\times10^{-1}$ and $5.6\times10^{-5}$. Here $N$ denotes the number of particles and $V$ denotes the system's volume. In what follows we will refer to the dimensionless density as simply the density. 

\subsection{The inverse power-law model}

The second model employed is also a 50:50 binary mixture of `large' and `small' particles, this time interacting via the pairwise potential
\begin{equation}
\varphi_{ij} =  \left\{\begin{array}{cc}\varepsilon_0\left(\left(\frac{r_{ij}}{l_{ij}}\right)^{-\beta} + \sum_{k=0}^{3}c_{2k}\left(\frac{r_{ij}}{l_{ij}}\right)^{2k}\right)\,,& \frac{r_{ij}}{l_{ij}} \le x_c\\0\,,&\frac{r_{ij}}{l_{ij}} > x_c\end{array}\right.,
\end{equation}
where $\varepsilon_0$ is a microscopic energy scale, and the dimensionless cutoff length was set to $x_c = 1.9$, which guarentees that the first coordination shell is always within the interaction range. The interaction lengths $l_{ij}$ were set to $l_0,1.18l_0$ and $1.4l_0$, respectively, where $l_0$ is a microscopic unit of length. The coefficients $c_{2k}$ are given by
\begin{equation}
c_{2k} = \frac{(-1)^{k+1}}{(6-2k)!!(2k)!!}\frac{(\beta+6)!!}{(\beta-2)!!(\beta+2k)}\cdot r_{c}^{-(\beta+2k)}\,,
\end{equation}
and ensure that the potential and 3 derivatives are continuous at $r_{ij}/l_{ij} = x_c$. We generated IPL solids under a fixed density of $\rho = 0.86$. 

The key control parameter in the IPL model is the exponent $\beta$, which we varied between 8 and 512. 

\subsection{Interaction cutoff}

In both models we introduce a cutoff in the pairwise potential for the sake of computational efficiency. This might appear to contradict the point of our work, which is to study the unjamming phenomena when such a cutoff is absent. We note here that the occurrence of unjamming phenomena in the canonical models begins to emerge when the \emph{first} coordination shell starts to approach the cutoff distance of the interactions. This is never the case in our systems, as we always set the interaction cutoff such that the first coordination shell is well within the interaction range. In other words, the first coordination shell never probes the cutoff distance in any meaningful way in our numerical experiments, therefore any unjamming phenomena we observe is independent of the existence of this cutoff. We have indeed verified that eliminating the cutoff altogether has a quantitatively negligible effect on our results. 

\subsection{Sample generation}

We created 1024 independent glassy samples of size $N=1600$ for both the IPL and EXP systems. We verified that finite size effects are negligible by simulating a systems of $N = 3249$ as well, but most data is reported for $N=1600$. For EXP we started by creating samples of density $\rho = 0.56$ by a quick quench to zero temperature from the melt, and generated lower density configurations by decreasing the density by factors of $10^{1/6}$, minimizing the solids after each such decrease. For densities lower than $2.6\times10^{-2}$, quad-precision numerics (i.e.~128 bit precision) were used. 

For IPL we chose $\beta = 12$ for the initial solid configurations, also generated by a quick quench from the melt. We then varied $\beta$ followed by an energy minimization to obtain glassy samples of other powers $\beta$. We employed quad-precision numerics for all IPL calculations. 

\subsection{Observables}

As commonly practiced in the field of unjamming \cite{liu_boson_peak_prl,vincenzo_epl_2010,brian_nonlocal_2016}, we calculated some of the observables in a shadow system for which the forces $-\frac{\partial \varphi}{\partial r}$ were set to zero. The shadow systems can be considered as relaxed elastic spring networks (i.e.~in which all springs reside at their respective rest-lengths) whose stiffnesses are given by the original pairwise potential stiffnesses $\frac{\partial^2\varphi}{\partial r^2}$. This procedure removes noise and the destabilizing effect of internal stresses, which has been shown to not affect scaling properties. The shadow system is referred to below as the `unstressed' system.  

\subsubsection{Elastic moduli}
Athermal elastic moduli were calculated following \cite{athermal_elasticity}. We used the definitions 
\begin{equation}
\mu \equiv \frac{1}{V}\frac{d^2U}{d\gamma^2}\quad\mbox{and}\quad B \equiv \frac{1}{V}\frac{d^2U}{\partial \eta^2}
\end{equation}
for the shear and bulk modulus, respectively, where $U$ is the potential energy, $\gamma$ is the simple shear strain and $\eta$ is the expansive strain. The latter parametrize the 2D strain tensor $\epsilon$ as follows
\begin{equation}
\epsilon = \frac{1}{2}\left( \begin{array}{cc}2\eta + \eta^2&\gamma + \gamma\eta\\\gamma + \gamma\eta&2\eta + \eta^2 +\gamma^2\end{array}\right)\,.
\end{equation}
In terms of the general first and second order moduli $C_{\kappa\chi} \equiv \frac{1}{V}\frac{dU}{d\epsilon_{\kappa\chi}}$ and $C_{\kappa\chi\theta\tau} \equiv \frac{1}{V}\frac{d^2U}{d\epsilon_{\kappa\chi}d\epsilon_{\theta\tau}}$ , our definitions of shear and bulk moduli read
\begin{equation}
\mu = C_{yy} + C_{xyxy}\,,
\end{equation}
and 
\begin{equation}
B = C_{xx} + C_{yy} + C_{xxxx} + C_{yyyy} + 2C_{xxyy}\,.
\end{equation}
We employed quad-precision numerics to calculate elastic moduli in all systems that were created using quad-precision. 

\subsubsection{Density of states}

We calculated the eigenvalues of the dynamical matrix ${\cal M}_{ij} = \frac{\partial^2U}{\partial\vec{R}_i\partial\vec{R}_j}$, where $\vec{R}_i$ denotes the $\dbar$ dimensional position vector of the $i^{\mbox{\tiny th}}$ particle, using standard open-source linear algebra libraries. The density of states $D(\omega)$ was obtained by histogramming over the square root of the eigenvalues, recalling that the masses are all unity.  

\subsubsection{Characteristic frequency scale}
We follow \cite{matthieu_le_epl} to probe a characteristic vibrational frequency scale in our glassy samples. This is done by considering the shadow relaxed spring system as described above, and imposing a unit dipolar force on the $i,j$ pair of the form
\begin{equation}\label{dipole}
\vec{d}^{\;ij}_k = (\delta_{jk} - \delta_{ik})\frac{\vec{R}_{ij}}{r_{ij}}\,.
\end{equation}
We calculate the responses
\begin{equation}
\vec{z}^{\;ij}_k =\tilde{\cal M}^{-1}_{km}\cdot\vec{d}^{\;ij}_m\,,
\end{equation}
where $\tilde{M}$ is the dynamical matrix of the shadow system. The characteristic frequency squared of the normalized responses $\hat{z}^{ij} = \vec{z}^{\;ij}/|\vec{z}^{\;ij}|$ are then calculated as
\begin{equation}
\omega_*^2 \equiv \overline{\hat{z}^{ij}_k\cdot\tilde{\cal M}_{km}\cdot\hat{z}^{ij}_m}\,,
\end{equation}
where $\overline{\circ}$ denotes an average over interacting pairs, and over our ensemble of glassy solid for each value of the control parameter. 

%%%%%%%%%%%%%%%%%%%%%%%%%%%%%%%%%%%%%%%%%%%%%%%%%%%%%%%
\begin{figure}[!ht]
\centering
\includegraphics[width = 0.49\textwidth]{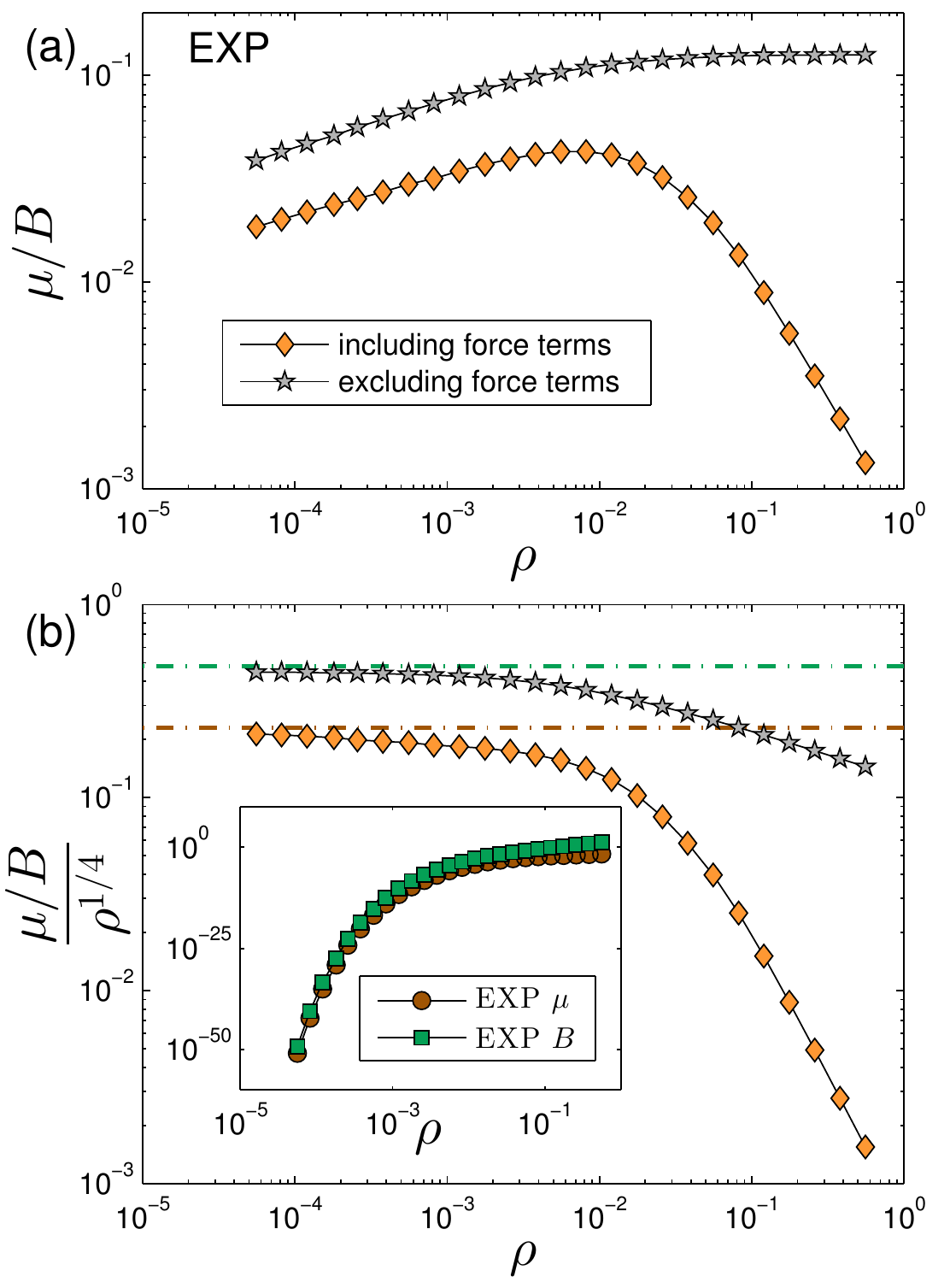}
\caption{\footnotesize a) Shear to bulk moduli ratio, measured as a function of density in the EXP system. In addition to the bare data represented by the orange diamonds, we also plot in gray stars the ratio calculated while omitting the terms that contain the interparticle forces, see text for further discussion. b) We find that the moduli data is consistent with our scaling argument (see Sect.~\ref{scaling_argument} which predicts $\mu/B \sim \rho^{1/4}$ (horizontal dash-dotted lines); while this scaling does not yet hold in the density regime accessible by our simulations, it is apparent that the curves are slowly converging to the predicted scaling at lower densities. The inset shows the bare moduli as a function of density. }
\label{exp_mu_over_B}
\end{figure}
%%%%%%%%%%%%%%%%%%%%%%%%%%%%%%%%%%%%%%%%%%%%%%%%%%%%%%

\section{Results}
\label{results}

\subsection{Shear to bulk moduli ratio}

In Fig.~\ref{exp_mu_over_B} we show our results for the shear and bulk moduli of the EXP system. The bare moduli are plotted vs.~density in the inset of panel \textbf{b}; we find that both moduli become exponentially small with with decreasing density, as expected from the form of the pairwise interaction potential. In panel \textbf{a} we plot the ratio of the shear to bulk moduli, which shows intriguing nonmonotonic behavior: as the density is decreased, we initially observe an \emph{increase} in $\mu/B$, up to a crossover density of approximately $\rho\approx10^{-2}$, which is further discussed in Sect.~\ref{scaling_argument}. Below this crossover, $\mu/B$ appears to vanish as $\rho \to 0$, which indicates the occurrence of an unjamming transition in the limit $\rho \to 0$. Omitting force terms in the calculation of the moduli causes the crossover to disappear altogether. The nonmonotonicity we find is reminiscent of the elastic behavior of highly compressed soft spheres as observed in \cite{massimo}.

In Sect.~\ref{scaling_argument} we argue that in the EXP model $\mu/B$ should scale as $\rho^{1/4}$ as $\rho \to 0$; we therefore plot in Fig.~\ref{exp_mu_over_B}b the rescaled ratio $\frac{\mu/B}{\rho^{1/4}}$. The dash-dotted lines are guides to the eye, showing that the measured data is consistent with our prediction, although we do not yet cleanly observe this scaling in the accessible density range.

%%%%%%%%%%%%%%%%%%%%%%%%%%%%%%%%%%%%%%%%%%%%%%%%%%%%%%%
\begin{figure}[!ht]
\centering
\includegraphics[width = 0.50\textwidth]{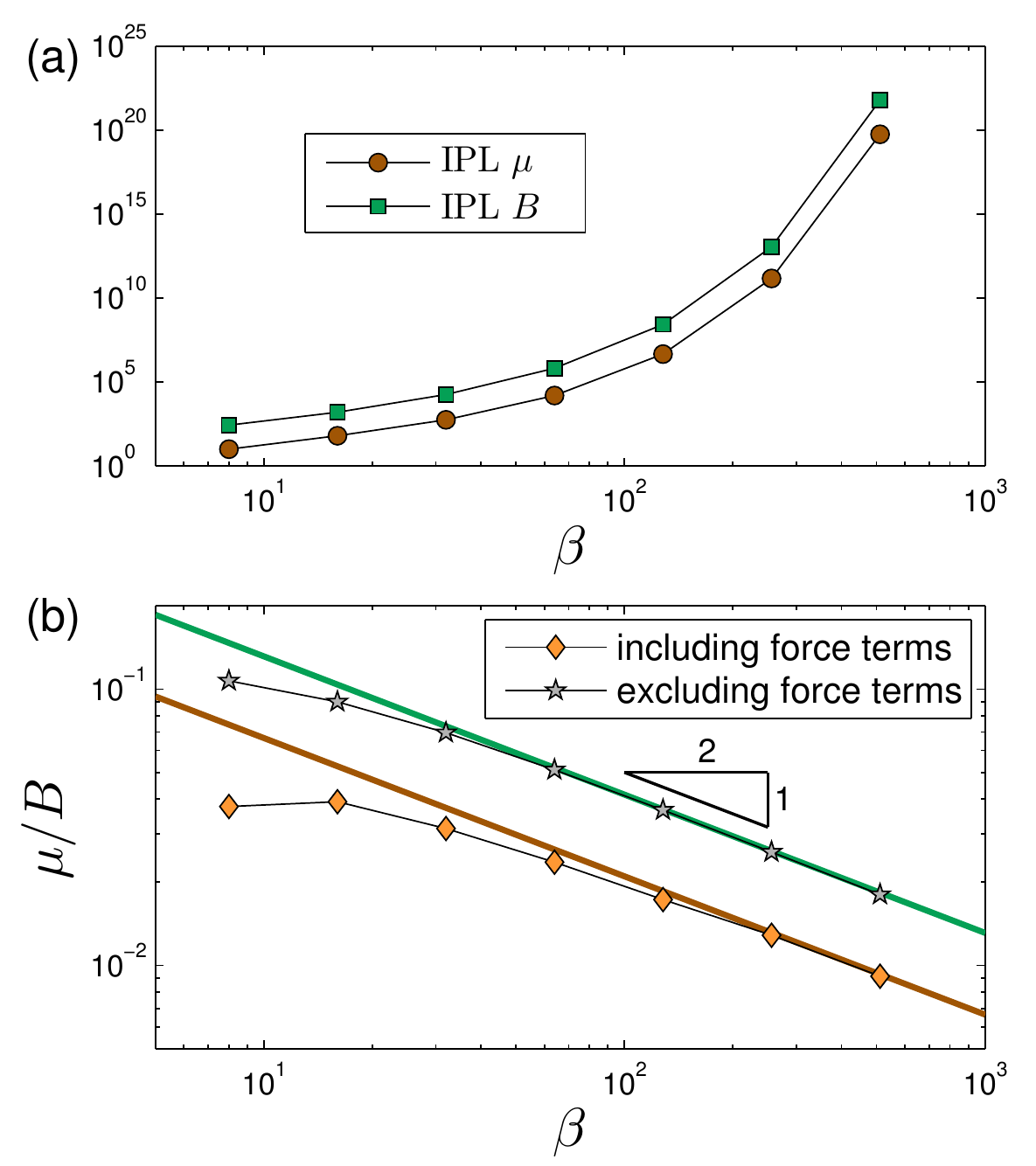}
\caption{\footnotesize a) Bare elastic moduli as a function of the exponent $\beta$, measured for the IPL samples. b) Shear to bulk moduli ratio for the stressed and unstressed IPL samples. The continuous lines represent the scaling $\mu/B \sim \beta^{-1/2}$, derived in Sect.~\ref{scaling_argument}.}
\label{ipl_mu_over_B}
\end{figure}
%%%%%%%%%%%%%%%%%%%%%%%%%%%%%%%%%%%%%%%%%%%%%%%%%%%%%%

In Fig.~\ref{ipl_mu_over_B} we display results for the shear and bulk moduli in the IPL system. Panel \textbf{a} shows the bare moduli, which appear to grow exponentially with increasing the exponent $\beta$ at fixed volume. Panel \textbf{b} shows the dependence of the ratio $\mu/B$ on the exponent $\beta$. Above a crossover at $\beta\! \approx\!200$, we find the scaling $\mu/B\! \sim\! \rho^{-1/2}$, as predicted for the IPL system in Sect.~\ref{scaling_argument}. We observe the same behavior for the shadow system, albeit with an earlier crossover at around $\beta \approx 30$. Our data indicates that in the IPL system unjamming occurs in the limit $\beta \to \infty$, where $\mu/B$ presumably vanishes. We note that varying $\beta$ by a factor of 4, between $\beta\!=\!8$ and $\beta\!=\!32$, the shear to bulk modulus ratio changes by merely 20\%, which is strong support of the quasi-universality of the IPL model put forward by Dyre and co-workers \cite{jeppe_nature_com,jeppe}, at least in the low-$\beta$ regime.

Interestingly, we find both in the EXP and IPL systems that the ratio of shear to bulk modulus is larger in the unstressed systems by a factor of $\approx 2$ upon approaching the unjamming point, precisely as predicted by Effective Medium Theory \cite{eric_emt_prestress}.

\subsection{Density of states}

\begin{figure}[!ht]
\centering
\includegraphics[width = 0.50\textwidth]{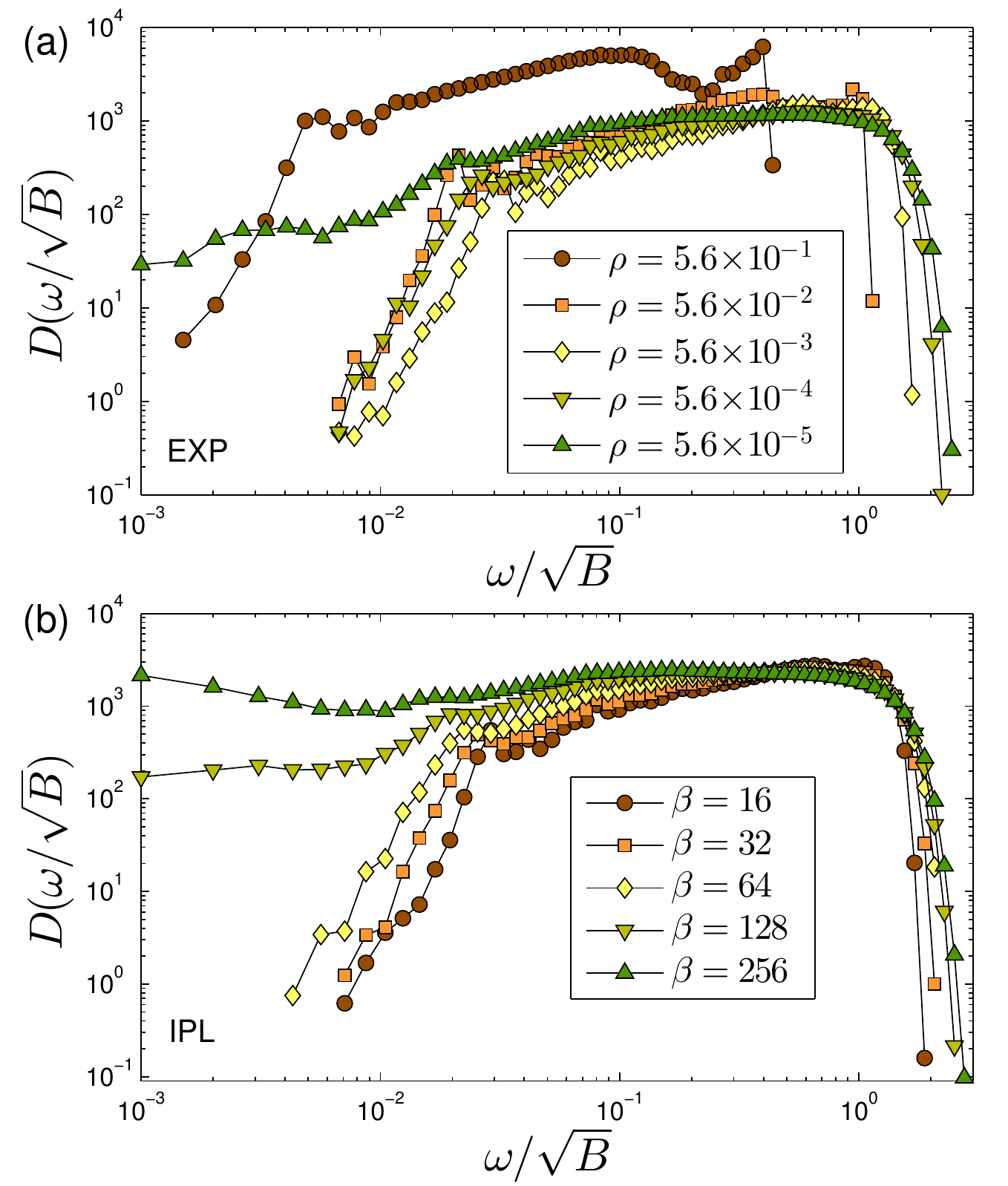}
\caption{\footnotesize (a) Density of states of the EXP system, at densities as indicated by the legend. (b) Density of states of the IPL system, calculated in states with exponents $\beta$ as indicated by the legend. We left out of this plot $D(\omega)$ obtained for the exponents $\beta\!=\!8$ and $\beta\!=\!512$ for visual clarity; we find that the same trend persists.}
\label{dos_fig}
\end{figure}

Another hallmark of unjamming is the appearance of an excess of low-frequency vibrational modes in the density of states (DOS) as the unjamming point is approached. Here we test whether and how this observation manifests itself in our EXP and IPL model systems. In Fig.~\ref{dos_fig} we plot the DOS averaged over our ensemble of glassy samples of our two models, as a function of the rescaled frequency $\omega/\sqrt{B}$, for values of the control parameter as indicated by the legends. We note that $\sqrt{B}$ (which has the required units of frequency in two dimensions, recalling that our units of mass $m\!=\!1$) is the natural \emph{high}-frequency scale in the unjamming problem. This is because the conventional Debye frequency is defined in terms of the shear modulus, which exhibits anomalies close to unjamming, and does not therefore well-represent the scale of high-frequency vibrational modes. 

%%%%%%%%%%%%%%%%%%%%%%%%%%%%%%%%%%%%%%%%%%%%%%%%%%%%%%%
%%%%%%%%%%%%%%%%%%%%%%%%%%%%%%%%%%%%%%%%%%%%%%%%%%%%%%

One generically expects the DOS to be supported by a larger and larger frequency range as the unjamming point is approached \cite{ohern2003,matthieu_thesis}. Our data for the DOS does not allow us to reliably extract a frequency scale that characterizes low-frequency modes. This point is further discussed in the next Subsection. We do, however, clearly see how the support of the DOS changes as the control parameter is varied.

The nonmonotonicity observed in $\mu/B$ for the EXP model is reflected by the unusual dependence of the DOS on density. For the highest density analyzed ($\rho\! = \!0.56$), the DOS exhibits an overall shift to low relative frequencies. At higher densities, we only see a clear increase in the support below $\rho\! =\! 5.6\!\times\!10^{-3}$, which becomes most pronounced at the lowest density analyzed, in which a clear excess of low-frequency modes appears. 

The IPL system shows a much clearer, monotonic increase in the support of the DOS as the exponent $\beta$ is increased. At the largest $\beta$ values analyzed ($\beta\!=\!128$ and $\beta\!=\!256$), a pronounced enhancement of the low-frequency tails of the DOS is observed.

\subsection{Characteristic frequency scale}

As mentioned in the previous Subsection, we are unable to reliably extract a characteristic low-frequency scale from our data of the DOS of both the EXP and IPL models. We resort therefore to extracting such a scale by different means; we follow \cite{matthieu_le_epl} and calculate `trial modes' as the (normalized) response to a local dipolar force applied on a pair of interacting particles, as explained in Sect.~\ref{numerics}. We chose to perform this calculation on the shadow unstressed system. 

%%%%%%%%%%%%%%%%%%%%%%%%%%%%%%%%%%%%%%%%%%%%%%%%%%%%%%%
\begin{figure}[!ht]
\centering
\includegraphics[width = 0.50\textwidth]{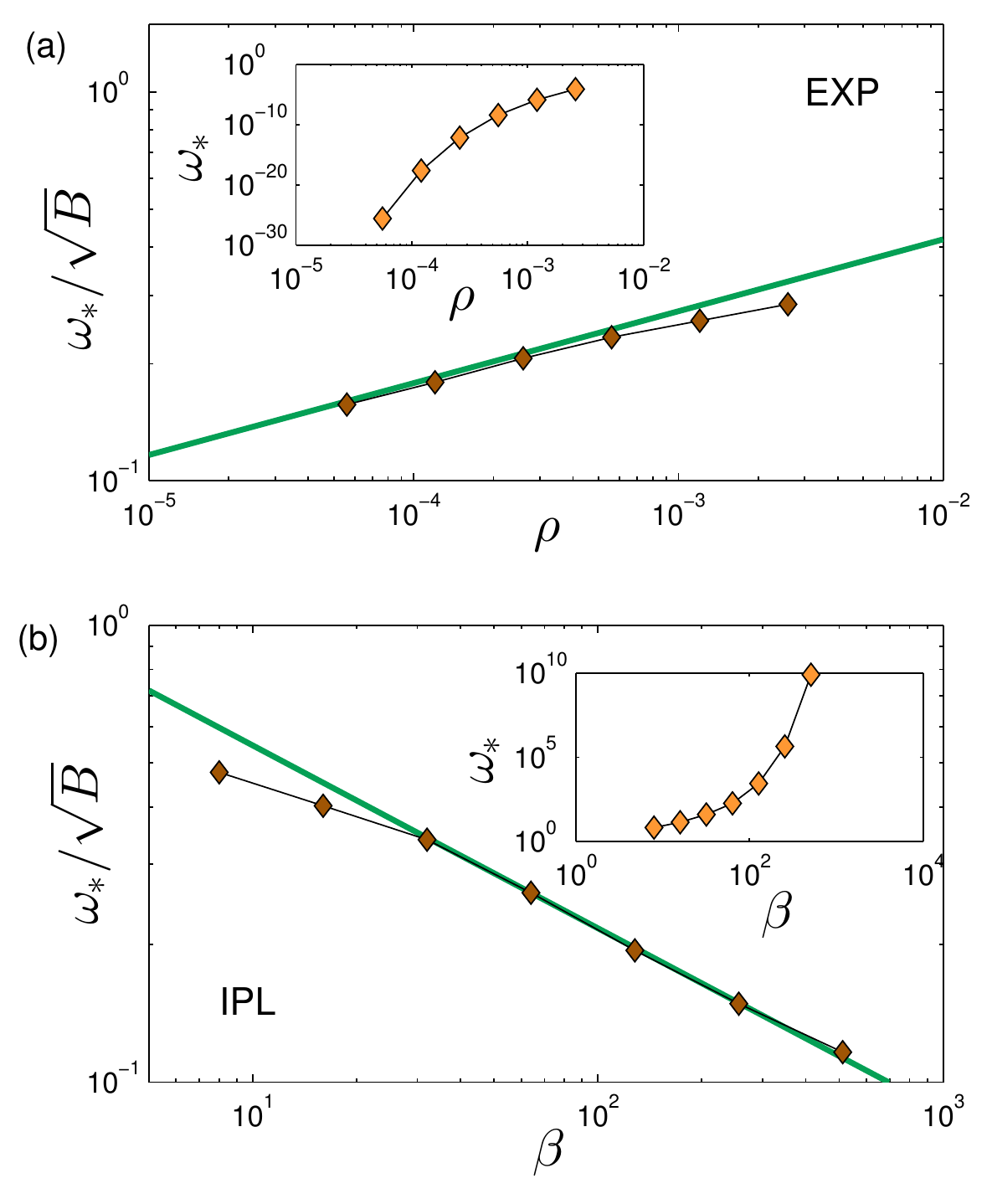}
\caption{\footnotesize Characteristic `unjamming' frequency scale $\omega_*$ expressed in terms of $\sqrt{B}$ for the EXP system (a) and the IPL system (b). Each data point represents the median calculated over 1600 responses. The insets display the bare medians of $\omega_*$ vs.~the control parameter. The continuous lines correspond to $\omega_*/\sqrt{B} \sim \rho^{0.19}$ and $\omega_*/\sqrt{B} \sim \beta^{-0.40}$ for the EXP and IPL systems, respectively.}
\label{freq_scale}
\end{figure}
%%%%%%%%%%%%%%%%%%%%%%%%%%%%%%%%%%%%%%%%%%%%%%%%%%%%%%

Fig.~\ref{freq_scale} displays our results; panel \textbf{a} shows the median of $\omega_*$ normalized by $\sqrt{B}$ for the EXP model, while the inset displays the bare medians of $\omega_*$.  We find that at low densities $\omega_*/\sqrt{B} \sim \rho^{0.19}$, which is represented by the continuous line. We similarly plot the rescaled median characteristic frequency $\omega_*/\sqrt{B}$ for the IPL in panel \textbf{b}, while the bare median characteristic frequency is shown in the inset. Here we find that at large exponents $\beta$, $\omega_*/\sqrt{B} \sim \beta^{-0.4}$. 

Interestingly, our scaling arguments spelled out in Sect.~\ref{scaling_argument} predict $\omega_*/\sqrt{B} \sim \rho^{1/4}$ for the EXP system, and $\omega_*/\sqrt{B} \sim \beta^{-1/2}$ for the IPL system. The exponents we measure are both smaller by approximately 20\% from the predicted ones. We attribute this disagreement to the imperfect correspondence between the bulk modulus and characteristic high vibrational frequency scales in our samples, as evident by the lack of collapse of the high-frequency tails of the DOS as shown in Fig.~\ref{dos_fig}.

\section{Discussion}
\label{scaling_argument}

We begin with discussing the relation between the hydrostatic pressure to bulk modulus ratio ($p/B$) and the distance to the unjamming point. To this aim we spell out the expressions for $p$ and $B$ in the athermal limit \cite{athermal_elasticity}, assuming the potential energy is expressed as a sum over radially-symmetric pairwise interactions:
\begin{eqnarray}
p & \equiv & -\frac{1}{V\dbar}\sum_{i<j}\varphi'_{ij}r_{ij} \,, \label{foo05} \\
B & \equiv & \frac{1}{V}\bigg( \sum_{i<j}\varphi''_{ij}r_{ij}^2 - \vec{\Xi}_k\cdot {\cal M}^{-1}_{k\ell}\cdot\vec{\Xi}_\ell\bigg) \label{foo00} \,, 
\end{eqnarray}
where $\vec{\Xi}_k \!\equiv\! \sum_{i<j}\varphi''_{ij}r_{ij}\vec{d}^{\;ij}_k$, and the dipole vector $\vec{d}^{\;ij}_k$ is defined in Eq.~(\ref{dipole}). Notice that $\vec{\Xi} = 0$ identically for the IPL model, which means that the second term on the RHS of Eq.~(\ref{foo00}), known as the `nonaffine' contribution to the bulk modulus, is identically zero in that system. 

\begin{figure}[!ht]
\centering
\includegraphics[width = 0.50\textwidth]{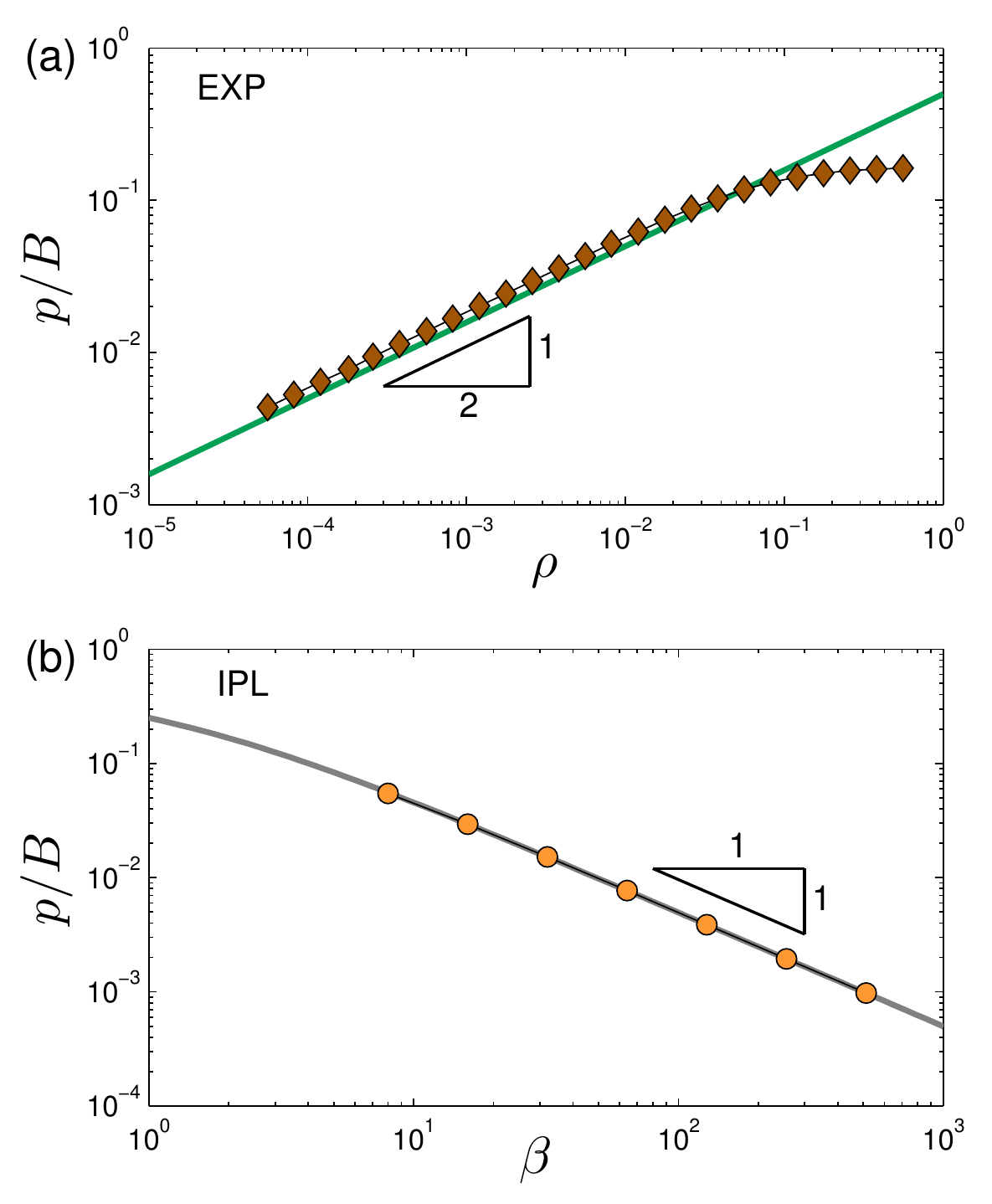}
\caption{\footnotesize Hydrostatic pressure to bulk modulus ratio for (a) the EXP and (b) the IPL systems, as a function of the relevant control parameter. The continuous line correspond the theoretical predictions Eqs.~(\ref{foo01}) and (\ref{foo02}) for the EXP and IPL systems, respectively. The prefactor of the scaling $p/B\!\sim\!\sqrt{\rho}$ for the EXP system is found to be about 0.58 instead of the predicted 1/2.}
\label{p_over_B_fig}
\end{figure}

Let us focus first on the EXP system, and express pairwise distances in terms of the density, namely $\tilde{r} \equiv r\sqrt{\rho}$. We now make the ansatz
\begin{equation}
N^{-1}\sum_{i<j}\varphi_{ij}'\tilde{r}_{ij} \approx N^{-1}\sum_{i<j}\varphi_{ij}''\tilde{r}_{ij}^2 \equiv g(\rho)\,,
\end{equation}
where $g(\rho)$ is an unknown function of the density. Using the ansatz in Eqs.~(\ref{foo05}) and (\ref{foo00}), recalling that $V \sim \rho^{-1}$, and neglecting for the moment the nonaffine term in Eq.~(\ref{foo00}), we write for the EXP system $p = \frac{\sqrt{\rho}}{2}g(\rho)$ and $B \approx g(\rho)$ (only valid in 2D, but with obvious generalization to 3D). From here we immediately see that
\begin{equation}\label{foo01}
p/B = \sqrt{\rho}/2\,,
\end{equation}
as verified in Fig.~\ref{p_over_B_fig}a, where it is shown that the scaling $p/B \sim \sqrt{\rho}$ is predicted perfectly, however the prefactor is off by roughly 15\% due to the approximation made in relating $p$ and $B$ to the ansatz function $g(\rho)$. 

%%%%%%%%%%%%%%%%%%%%%%%%%%%%%%%%%%%%%%%%%%%%%%%%%%%%%%%
%%%%%%%%%%%%%%%%%%%%%%%%%%%%%%%%%%%%%%%%%%%%%%%%%%%%%%

We learn from the good agreement of Eq.~(\ref{foo01}) with our numerics that neglecting the nonaffine contribution to the bulk modulus is a reasonable approximation close to unjamming. This can be justified as follows: compare the vectors $\vec{\Xi}_k \!=\! \sum_{i<j}\varphi''_{ij}r_{ij}\vec{d}^{\;ij}_k$ and the net forces $\vec{F}_k\!=\! -\sum_{i<j}\varphi'_{ij}\vec{d}^{\;ij}_k$; the latter are identically zero due to mechanical equilibrium. Considering that in systems of purely repulsive interactions stiffnesses and forces are correlated (and more at low densities in the EXP model), one would indeed expect that the vector $\vec{\Xi}$ would also be small in magnitude, resulting in a negligible nonaffine contribution to the bulk modulus.

The situation is more straightforward for the IPL system, where the nonaffine contribution to the bulk modulus vanishes identically. Here we make the ansatz
\begin{equation}
V^{-1}\sum_{i<j}r_{ij}^{-\beta} \equiv f(\beta)\,,
\end{equation}
where $f(\beta)$ is an unknown function of the exponent $\beta$. Using this ansatz, we write for the IPL system $p = \frac{\beta}{2} f(\beta)$ and $B = \beta(\beta+1)f(\beta)$, then we expect
\begin{equation}\label{foo02}
p/B = \frac{1}{2(\beta+1)}\,,
\end{equation}
as verified in Fig.~\ref{p_over_B_fig}b.

\begin{figure}[!ht]
\centering
\includegraphics[width = 0.50\textwidth]{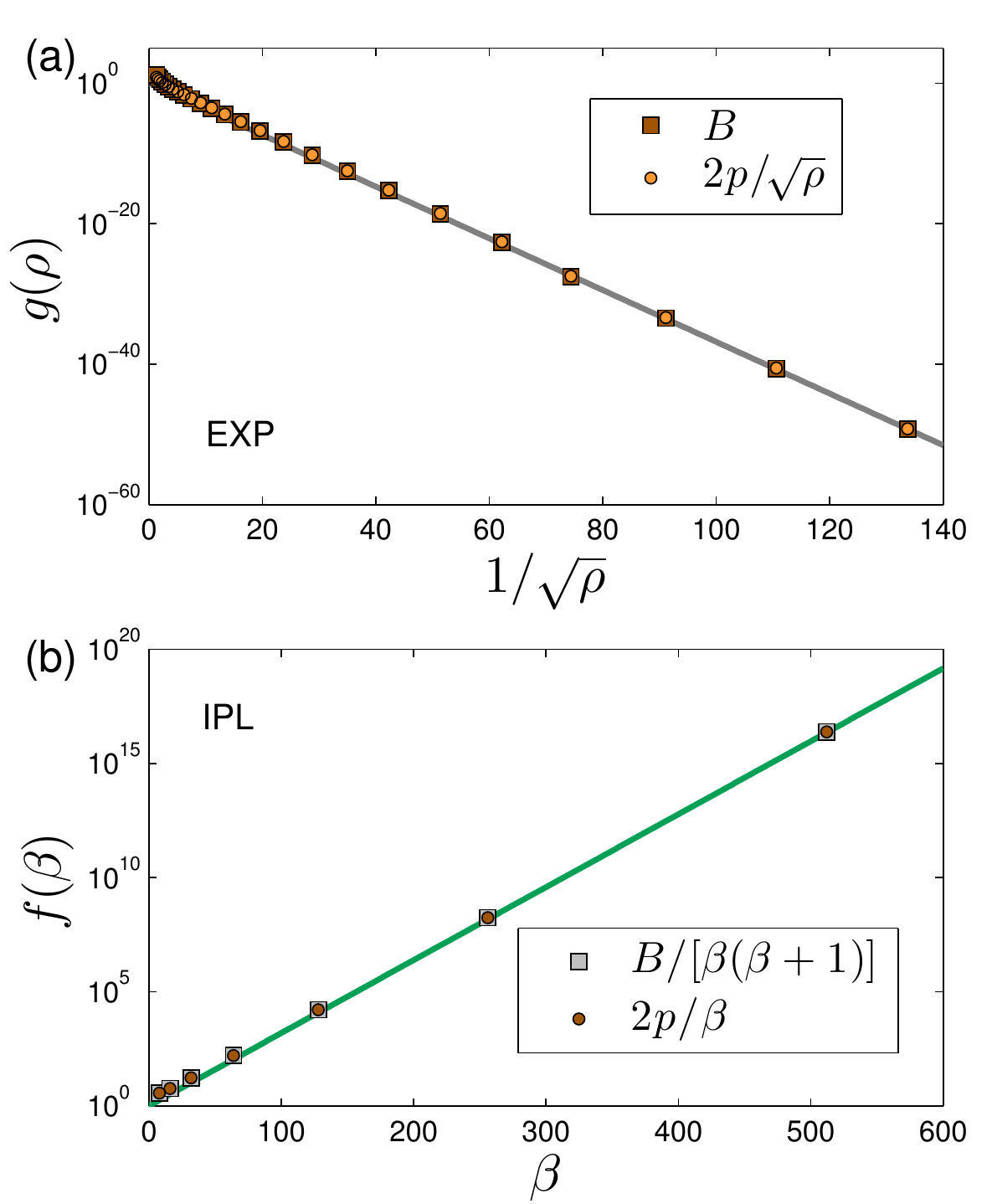}
\caption{\footnotesize Ansatz functions $g(\rho)$ (a) and $f(\beta)$ (b), calculated as explained in the text using both the hydrostatic pressure and bulk modulus data. The continuous lines represent Eqs.~(\ref{foo03}) and (\ref{foo04}).}
\label{ansatz_fig}
\end{figure}

In Fig.~\ref{ansatz_fig} we plot the ansatz functions $g(\rho)$ and $f(\beta)$, calculated using both the pressure and bulk modulus data for the EXP and IPL systems. As unjamming is approached, we find very good agreement between the two calculations for both functions, which are empirically found to fit very well the following functional forms:
\begin{eqnarray}
g(\rho) & = & e^{-\sqrt{0.72/\rho}}\,, \label{foo03} \\
f(\beta) & = & e^{\beta/13.6}\,, \label{foo04}
\end{eqnarray}
which are represented by the continuous lines in Fig.~\ref{ansatz_fig}a and Fig.~\ref{ansatz_fig}b, respectively.

%%%%%%%%%%%%%%%%%%%%%%%%%%%%%%%%%%%%%%%%%%%%%%%%%%%%%%%
%%%%%%%%%%%%%%%%%%%%%%%%%%%%%%%%%%%%%%%%%%%%%%%%%%%%%%

The exponential dependence of $g(\rho)$ on $1/\sqrt{\rho}$ arises naturally from the form of the interaction potential of the EXP system. The density scale $\rho_0 \!\approx\! 0.72$ is also consistent with the interaction length parameters $l_{ij}$, which were chosen to be equal or slightly larger than unity. We emphasize that the argumentation spelled out above is dimension dependent, and here we only focus on 2D. 

The exponential form of $f(\beta)$ can be understood by differentiating the pressure or bulk modulus with respect to $\beta$; one finds then that the $f(\beta)$ should crucially depend on the density considered: for instance, at large densities one expects the bulk modulus and pressure to grow with increasing $\beta$, whereas for small densities the opposite behavior should occur. The scale that describes the exponential increase in $f(\beta)$, found to be approximately 14 in our system, is related to the (logarithm of) the characteristic ratio between typical pairwise distances, and the interaction length parameters $l_{ij}$. 

One well-known result from the unjamming literature \cite{ohern2003,matthieu_thesis} relates the coordination difference to the isostatic point $\delta z \!\equiv\! z-2\dbar$ to the pressure to bulk modulus ratio as $\delta z \!\sim\! \sqrt{p/B}$. We can use this relation to define an effective coordination in our systems (which lack a clear-cut definition of connectivity). For example, the canonical KABLJ system \cite{kablj}, in which pairwise interactions can be effectively described by a $r^{-18}$ law to a good approximation \cite{jeppe_2008}, would be assigned an effective $\delta z$ of order unity.

We can further use the previously established results from the unjamming literature \cite{ohern2003,matthieu_thesis}: $\mu/B\!\sim\!\delta z \!\sim\! \sqrt{p/B}$ and $\omega_*/\sqrt{B} \!\sim\! \delta z \!\sim\! \sqrt{p/B}$, to predict the dependence of the shear to bulk modulus ratio and the characteristic frequency scale on the distance to unjamming in our model systems. In particular we expect
\begin{eqnarray}
\mu/B\sim\omega_*/\sqrt{B}\sim \rho^{1/4} \quad \mbox{in the EXP system}\,, \\
\mu/B\sim\omega_*/\sqrt{B}\sim \beta^{-1/2} \quad \mbox{in the IPL system}\,,
\end{eqnarray}
in good agreement with our numerical results for $\mu/B$ displayed in Figs.~\ref{exp_mu_over_B} and \ref{ipl_mu_over_B}, and in reasonable consistency with our numerical results for $\omega_*/\sqrt{B}$ displayed in Fig.~\ref{freq_scale}.

\section{Summary}
\label{summary}

In this work we have studied the unjamming behavior of two computer model glass forming systems of purely repulsive particles that interact via pairwise potentials with no sharp cutoffs in their respective interaction range. These models differ significantly from the canonical unjamming models, in which the sharp cutoff of the interaction range gives rise to unjamming once this cutoff probes the characteristic size of the first coordination shell of a particle (conventionally achieved by reducing the packing fraction or density). Despite the absence of a sharp cutoff in our models, we are still able to observe the hallmark phenomenology associated to the unjamming transition, in particular the vanishing of the shear to bulk moduli ratio, the emergence of excess low-frequency vibrational modes, and the vanishing of a characteristic frequency scale.

In the EXP model unjamming occurs in the limit $\rho \to 0$, and not at a finite density as in the canonical models. We find that the shear to bulk modulus ratio vanishes in good agreement with our scaling argument, which predicts $\mu/B \sim \rho^{1/4}$. We also find that the characteristic frequency scale $\omega_*/\sqrt{B}$ vanishes upon unjamming as $\rho^{0.19}$, which is close to, but not in perfect agreement with, our prediction $\omega_*/\sqrt{B} \sim \rho^{1/4}$. 

In the IPL model unjamming occurs in the limit $\beta \to \infty$: we find that the shear to bulk modulus ratio vanishes as $\mu/B \sim \beta^{-1/2}$, in good agreement with our theoretical prediction. The characteristic frequency scale is found to follow $\omega_*/\sqrt{B} \sim \beta^{-0.40}$, not far from our prediction $\omega_*/\sqrt{B} \sim \beta^{-1/2}$.

Our predictions are based on a simple ansatz used to find the relation between the pressure to bulk modulus ratio and the distance to unjamming. Using previously established result, this allows us to assign an effective excess  coordination $\delta z$ to each of our model systems, in which the connectivity cannot be cleanly defined. Once the dependence of the pressure to bulk modulus ratio is established, we use well-known results from the unjamming literature to predict the dependence of the shear to bulk modulus ratio and of the characteristic frequency scale (expressed in terms of the bulk modulus) on the distance to unjamming.

Our work highlights the importance of the pressure to bulk modulus ratio as a key dimensionless number that quantifies the distance to the unjamming point of any system with purely repulsive interactions, and the generality of the irrelevance of nonaffine contributions to the bulk modulus in such systems. 

\acknowledgements
We warmly thank Gustavo D\"uring for fruitful discussions. We would also like to thank SURFsara for the support in using the Lisa Computer Cluster.

\bibliography{references_lerner}

\end{document}